\documentstyle[12pt]{article}
\input psfig
\newcommand{\bea}{\begin{eqnarray}}
\newcommand{\eea}{\end{eqnarray}}
\newcommand{\beq}{\begin{equation}}
\newcommand{\eeq}{\end{equation}}

\def\msbar{\ifmmode{\overline{\rm MS}} \else{$\overline{\rm MS}$} \fi}
\def\drbar{\ifmmode{\overline{\rm DR}} \else{$\overline{\rm DR}$} \fi}
\def\st{\ifmmode{\tilde{t}} \else{$\tilde{t}$} \fi}
\def\sb{\ifmmode{\tilde{b}} \else{$\tilde{b}$} \fi}
\def\sq{\ifmmode{\tilde{q}} \else{$\tilde{q}$} \fi}
\def\sg{\ifmmode{\tilde{g}} \else{$\tilde{g}$} \fi}

\begin{document}
\pagestyle{empty}
\vspace*{-3cm}
\begin{flushright}
UWThPh-1995-26\\
HEPHY-PUB 626/95\\
TGU-15\\
ITP-SU-95/02\\
KEK-TH-447\\
hep-ph/9508283\\
\vspace{0.3cm}
August, 1995
\end{flushright}

\vspace{1cm}
\begin{center}
\begin{Large} \bf
QCD corrections to the decay $H^+ \rightarrow \st \bar{\sb}$\\
in the Minimal Supersymmetric \\
Standard Model
\end{Large}
\end{center}
\vspace{10mm}

\begin{center}
\large A.~Bartl,$^1$ H.~Eberl,$^2$ K.~Hidaka,$^3$ T.~Kon,$^4$\\
W.~Majerotto$^2$ and Y.~Yamada$^5$
\end{center}
\vspace{0mm}

\begin{center}
\begin{tabular}{l}
$^1${\it Institut f\"ur Theoretische Physik, Universit\"at Wien, A-1090
Vienna, Austria}\\
$^2${\it Institut f\"ur Hochenergiephysik der \"Osterreichische Akademie
der Wissenschaften,}\\
{\it A-1050 Vienna, Austria}\\
$^3${\it Department of Physics, Tokyo Gakugei University, Koganei,
Tokyo 184, Japan}\\
$^4${\it Faculty of Engineering, Seikei University, Musashino, Tokyo 180,
Japan}\\
$^5${\it Theory Group, National Laboratory for High Energy Physics (KEK)},\\
{\it Tsukuba, Ibaraki 305, Japan}\\
\end{tabular}
\end{center}
\vspace{10mm}

\begin{abstract}
\begin{small}
\baselineskip=28pt
We calculate the supersymmetric ${\cal O}(\alpha_s)$ QCD corrections to the
width of the decay
$H^+ \rightarrow \st\bar{\sb}$ within the Minimal Supersymmetric Standard
Model. We find that
the QCD corrections are significant, but that they do not invalidate our
previous conclusion
at tree-level on the dominance of the $\st\bar{\sb}$ mode in a wide
parameter region.
\end{small}
\end{abstract}

%\vfill

\newpage
\pagestyle{plain}
\setcounter{page}{2}
\baselineskip=28pt

\section{Introduction}
In the Minimal Supersymmetric Standard Model (MSSM) \cite{1,2} two Higgs
doublets are necessary,
leading to five physical Higgs bosons $h^0, H^0, A^0$, and $H^\pm$
\cite{3,4}. If all supersymmetric
(SUSY) particles are very heavy, the charged Higgs boson $H^+$ decays
dominantly into $t \bar b$; the
decays $H^+ \rightarrow \tau^+ \nu$ and/or $H^+ \rightarrow W^+ h^0$ are
dominant below the $t \bar b$
threshold \cite{3,4b}. In ref.~\cite{5} all decay modes of $H^+$ including
the SUSY-particle modes were
studied in detail; it was shown that the SUSY decay modes $H^+ \rightarrow
\st_i \bar{\sb}_j (i,j = 1,2)$
can be dominant in a large region of the MSSM parameter space due to large
$t$ and $b$ quark Yukawa
couplings and large $\st$- and $\sb$-mixings, and that this could have a
decisive impact on $H^+$
searches at future colliders.
Here $\st_i \, (\sb_j)$ are the scalar top (scalar bottom) mass eigenstates
which are mixtures of
$\st_L$ and $\st_R$ ($\sb_L$ and $\sb_R$).

The standard QCD corrections are very large for the width of $H^+
\rightarrow c \bar s$ and can be
large ($+10$\% to $-50$\%) for that of $H^+\rightarrow t \bar b$ \cite{6a}.
The QCD corrections from the SUSY-particle loops are calculated within the
MSSM for $H^+\rightarrow
t \bar b$ in \cite{6b} and turn out to be non-negligible ($\sim$ 10\%)
for certain values of the MSSM parameters. This suggests that the QCD
corrections to
$H^+\rightarrow\st\bar{\sb}$ could
also be large. Therefore it should be examined whether the result in
\cite{5} remains valid after
including the QCD corrections.

In this paper we calculate the ${\cal O}(\alpha_s)$ QCD corrections to the
width of $H^+ \rightarrow
\st_i \bar{\sb}_j$ within the MSSM. To the best of our knowledge they are
not known in the literature.
We obtain the complete ${\cal O}(\alpha_s)$ corrected width in the \drbar
renormalization scheme (i.e.
the \msbar scheme with dimensional reduction \cite{7}) including all quark
mass terms and $\sq_L-\sq_R$ mixings.
The main complication here is that the $\sq_L-\sq_R$ mixing angles are
renormalized by the SUSY QCD
corrections. We find that the corrections to the $\st\bar{\sb}$ width are
significant but that the
$\st\bar{\sb}$ mode is still dominant in a wide parameter range.

\section{Tree level result}
We first review the tree level results \cite{5}. The squark mass matrix in
the basis ($\sq_L$, $\sq_R$),
with $\sq=\st$ or $\sb$, is given by \cite{3,4} \beq
\left( \begin{array}{cc}m_{LL}^2 & m_{LR}^2 \\ m_{RL}^2 & m_{RR}^2
\end{array} \right)=
(R^{\sq})^{\dagger}\left( \begin{array}{cc}m_{\sq_1}^2 & 0 \\ 0 &
m_{\sq_2}^2 \end{array}
\right)R^{\sq}, \eeq
where
\bea
m_{LL}^2 &=& M_{\tilde{Q}}^2+m_q^2+m_Z^2\cos 2\beta (I_q-Q_q\sin^2\theta_W), \\
m_{RR}^2 &=& M_{\{\tilde{U},
\tilde{D}\}}^2+m_q^2+m_Z^2\cos 2\beta Q_q\sin^2
\theta_W, \\
m_{LR}^2=m_{RL}^2 &=& \left\{ \begin{array}{ll}
m_t(A_t-\mu\cot\beta) & (\sq=\st) \\
m_b(A_b-\mu\tan\beta) & (\sq=\sb) \end{array} \right. , \eea
and
\beq \label{5}
R^{\sq}_{i\alpha}=\left(
\begin{array}{cc}\cos\theta_\sq & \sin\theta_\sq \\
-\sin\theta_\sq & \cos\theta_\sq \end{array}\right) . \eeq
Here the mass eigenstates $\sq_i(i=1,2)$ (with $m_{\sq_1}<m_{\sq_2}$) are
related to the
SU(2)$_L$ eigenstates $\sq_{\alpha}(\alpha=L,R)$ as
$\sq_i=R^{\sq}_{i\alpha}\sq_{\alpha}$. Note that
in the sign convention used here the parameters $A_{t,b}$ correspond to
$(-A_{t,b})$ of ref.\cite{5}.

The tree-level decay width of $H^+ \rightarrow \st_i \bar{\sb}_j$ is then
given by (see Fig.~1a)
\beq \label{6}
\Gamma^{(0)}(H^+\rightarrow\st_i\bar{\sb}_j) =\frac{N_C\kappa}{16\pi
m_H^3}|G_{ij}|^2\,, \eeq
where $m_H$ is the $H^+$ mass,
$\kappa=\kappa(m_H^2, m_{\st_i}^2, m_{\sb_j}^2 )$, $\kappa(x,y,z)\equiv
((x-y-z)^2-4yz)^{1/2}$, $N_C=3$,
and \bea \label{7}
G_{ij}&=& \frac{g}{\sqrt{2}m_W}
R^{\st}\left( \begin{array}{cc}
m_b^2\tan\beta+m_t^2\cot\beta-m_W^2\sin 2\beta & m_b(A_b\tan\beta+\mu) \\
m_t(A_t\cot\beta+\mu) &
2m_tm_b/\sin 2\beta \end{array} \right)(R^{\sb})^{\dagger} \nonumber\\ \eea
are the $H^+\bar{\st}_i\sb_j$ couplings \cite{3,4}, with $g$ being the
SU(2) coupling.

\section{QCD virtual corrections}
The ${\cal O}(\alpha_s)$ QCD virtual corrections to
$H^+\rightarrow\st_i\bar{\sb}_j$ stem from the
diagrams of Fig.~1b (vertex corrections) and 1c (wave function
corrections). For simplicity we use
in this paper the \drbar renormalization scheme\footnote{Strictly speaking,
our renormalization
scheme is the $\drbar'$ scheme \cite{7b} where the ``$\epsilon$-scalar
mass'' is absorbed into
$M_{\tilde{Q},\tilde{U},\tilde{D}}^2$.} for all parameters which receive
the QCD corrections, i.e.
$m_{t,b}$, $A_{t,b}$, and $M_{\tilde{Q},\tilde{U},\tilde{D}}$. The
renormalized squark
mixing angle $\theta_{\sq}$ is then defined by the relations (1--5) in
terms of the \drbar parameters
$m_{t,b}$, $A_{t,b}$, and $M_{\tilde{Q},\tilde{U},\tilde{D}}$.

The one-loop corrected decay amplitudes $G_{ij}^{\rm corr}$ are expressed
as \beq
G_{ij}^{\rm corr}=G_{ij}+\delta G_{ij}^{(v)}+\delta G_{ij}^{(w)}, \eeq
where $G_{ij}$ are defined by (7) in terms of the \drbar parameters, and
$\delta G_{ij}^{(v)}$ and
$\delta G_{ij}^{(w)}$ are the vertex and squark wave function corrections,
respectively.

The vertex corrections $\delta G_{ij}^{(v)}$ are calculated from the graphs
of Fig.~1b as
\bea
\delta G_{ij}^{(v)}&=& \frac{\alpha_s C_F}{4\pi} \left[ \{B_0(m_{\st_i}^2,
0, m_{\st_i}^2)+B_0
(m_{\sb_j}^2, 0, m_{\sb_j}^2) -B_0(m_H^2, m_{\st_i}^2, m_{\sb_j}^2)
\right. \nonumber \\
&& -2(m_H^2-m_{\st_i}^2-m_{\sb_j}^2)
C_0(m_{\sb_j}^2, \lambda^2, m_{\st_i}^2)\}G_{ij} \nonumber \\ &&
-B_0(m_H^2, m_{\st_k}^2,
m_{\sb_l}^2)G_{kl}S^{\st}_{ik}S^{\sb}_{lj} \nonumber \\
&& +2\{ (\alpha_{LL})_{ij}(m_ty_2+m_by_1)+(\alpha_{RR})_{ij}(m_ty_1+m_by_2)\}
B_0(m_H^2, m_b^2, m_t^2) \nonumber\\
&& +2\{ (\alpha_{LL})_{ij}m_ty_2+(\alpha_{LR})_{ij}m_{\sg}y_1
+(\alpha_{RL})_{ij}m_{\sg}y_2+
(\alpha_{RR})_{ij}m_ty_1\} B_0(m_{\st_i}^2, m_{\sg}^2, m_t^2) \nonumber\\ && +
2\{ (\alpha_{LL})_{ij}m_by_1+(\alpha_{LR})_{ij}m_{\sg}y_1
+(\alpha_{RL})_{ij}m_{\sg}y_2+
(\alpha_{RR})_{ij}m_by_2\} B_0(m_{\sb_j}^2, m_{\sg}^2, m_b^2) \nonumber\\ &&
+2\{ (m_t^2+m_b^2-m_H^2)m_{\sg}
((\alpha_{LR})_{ij}y_1+(\alpha_{RL})_{ij}y_2) \nonumber\\ &&
+(m_b^2+m_{\sg}^2-m_{\sb_j}^2)m_t
((\alpha_{LL})_{ij}y_2+(\alpha_{RR})_{ij}y_1) \nonumber\\ &&
+(m_t^2+m_{\sg}^2-m_{\st_i}^2)m_b
((\alpha_{LL})_{ij}y_1+(\alpha_{RR})_{ij}y_2) \nonumber\\ && \left. +
2m_{\sg}m_tm_b((\alpha_{LR})_{ij}y_2+(\alpha_{RL})_{ij}y_1) \} C_0(m_b^2,
m_{\sg}^2, m_t^2) \right] ,
\eea
where $C_F=4/3$,
\beq
S^{\sq}=\left( \begin{array}{cc} \cos 2\theta_\sq & -\sin 2\theta_\sq \\
-\sin 2\theta_\sq & -\cos 2\theta_\sq
\end{array} \right) ,
\eeq
\[
\alpha_{LL}=\left( \begin{array}{rr}
\cos\theta_\st\cos\theta_\sb & -\cos\theta_\st\sin\theta_\sb \\
-\sin\theta_\st\cos\theta_\sb &
\sin\theta_\st\sin\theta_\sb \end{array} \right) , \;\;\;
\alpha_{LR}=\left( \begin{array}{rr}
-\cos\theta_\st\sin\theta_\sb & -\cos\theta_\st\cos\theta_\sb \\
\sin\theta_\st\sin\theta_\sb &
\sin\theta_\st\cos\theta_\sb \end{array} \right) , \]
\beq
\alpha_{RL}=\left( \begin{array}{rr}
-\sin\theta_\st\cos\theta_\sb & \sin\theta_\st\sin\theta_\sb \\
-\cos\theta_\st\cos\theta_\sb &
\cos\theta_\st\sin\theta_\sb \end{array} \right) , \;\;\;
\alpha_{RR}=\left( \begin{array}{rr}
\sin\theta_\st\sin\theta_\sb & \sin\theta_\st\cos\theta_\sb \\
\cos\theta_\st\sin\theta_\sb &
\cos\theta_\st\cos\theta_\sb \end{array} \right) ,
\eeq
\beq
y_1=\frac{g}{\sqrt{2}m_W}m_b\tan\beta=h_b\sin\beta, \;\;\;
y_2=\frac{g}{\sqrt{2}m_W}
m_t\cot\beta=h_t\cos\beta, \eeq
and $m_{\sg}$ is the gluino mass. A gluon mass $\lambda$ is introduced to
regularize
the infrared divergences.
Here we define the functions $A$, $B_0$, $B_1$ and $C_0$ as in \cite{pv}
($\Delta = 2/(4-D) - \gamma_E +\log 4\pi$): \bea
A(m^2)&=&
\int\frac{d^Dq}{i\pi^2}\frac{1}{q^2-m^2}=m^2(\Delta +\log(Q^2/m^2)+1), \\
B_0(k^2, m_1^2, m_2^2)&=&
\int\frac{d^Dq}{i\pi^2}\frac{1}{(q^2-m_1^2)((q+k)^2-m_2^2)} \nonumber\\
&=&\Delta-\int_0^1dz\log\frac{(1-z)m_1^2+zm_2^2-z(1-z)k^2-i\delta}{Q^2}, \\
B_1(k^2, m_1^2, m_2^2)&=&
\frac{1}{k_{\mu}}\int\frac{d^Dq}{i\pi^2}
\frac{q_{\mu}}{(q^2-m_1^2)((q+k)^2-m_2^2)} \nonumber\\
&=&-\frac{\Delta}{2}+\int_0^1dz\,z\log
\frac{(1-z)m_1^2+zm_2^2-z(1-z)k^2-i\delta}{Q^2}, \\ C_0(m_1^2, m_2^2, m_3^2)&=&
\int\frac{d^Dq}{i\pi^2}\frac{1}{(q^2-m_1^2)((q+k_2)^2-m_2^2)
((q+p)^2-m_3^2)} \nonumber\\
&=&
-\int_0^1dx\int_0^1dy\int_0^1dz\delta(1-x-y-z)\,
\times \nonumber\\
&&(xm_1^2+ym_2^2+zm_3^2-xym_{\sb_j}^2
-yzm_{\st_i}^2-xzm_H^2-i\delta)^{-1}.
\eea
Here $p$ and $k_2$ are respectively the external momenta of $H^+$ and
$\bar{\sb}_j$, and
$Q$ is the \drbar renormalization scale. Note that $\Delta$ is omitted in
the \drbar scheme.

The squark wave function corrections $\delta G_{ij}^{(w)}$ are expressed as
\beq \label{17}
\delta G_{ij}^{(w)}=-\frac{1}{2}\left[ \dot{\Pi}_{ii}^{\st}(m_{\st_i}^2)
+\dot{\Pi}_{jj}^{\sb}(m_{\sb_j}^2)\right] G_{ij}
-\frac{\Pi_{ii'}^{\st}(m_{\st_i}^2)}{m_{\st_i}^2-m_{\st_{i'}}^2} G_{i'j}
-\frac{\Pi_{j'j}^{\sb}(m_{\sb_j}^2)}{m_{\sb_j}^2-m_{\sb_{j'}}^2} G_{ij'},
\eeq
where $i\neq i'$ and $j\neq j'$.
$\Pi_{ij}^{\sq}(k^2)$ are the one-loop corrections to the two-point
functions of
$\bar{\sq}_i\sq_j$, which are obtained from the graphs of Fig.~1c.
$\dot{\Pi}(k^2)$
denotes the derivative with respect to $k^2$. The last two terms in (\ref{17})
represent the corrections due to the renormalization of the $\sq$-mixings.
The explicit forms are
\bea
\dot{\Pi}_{ii}^{\sq}(m_{\sq_i}^2)&=&
\frac{\alpha_sC_F}{4\pi}\left[ -3B_0(m_{\sq_i}^2, 0, m_{\sq_i}^2)
-2B_1(m_{\sq_i}^2, 0, m_{\sq_i}^2)
-4m_{\sq_i}^2\dot{B}_0(m_{\sq_i}^2, \lambda^2, m_{\sq_i}^2) \right. \nonumber\\
&&-2m_{\sq_i}^2\dot{B}_1(m_{\sq_i}^2, 0, m_{\sq_i}^2)
-4m_{\sg}^2\dot{B}_0(m_{\sq_i}^2, m_{\sg}^2, m_q^2) -4B_1(m_{\sq_i}^2,
m_{\sg}^2, m_q^2) \nonumber\\
&&\left. -4m_{\sq_i}^2\dot{B}_1(m_{\sq_i}^2, m_{\sg}^2, m_q^2)
+(-)^{i-1}4\sin 2\theta_\sq m_qm_{\sg}\dot{B}_0(m_{\sq_i}^2, m_{\sg}^2,
m_q^2) \right] ,
\eea
and
\beq
\Pi_{i'i}^{\sq}(m_{\sq_i}^2)=\frac{\alpha_sC_F}{4\pi}
\left[ \frac{1}{2}\sin 4\theta_\sq (A(m_{\sq_2}^2)-A(m_{\sq_1}^2))
+4\cos 2\theta_\sq m_qm_{\sg}B_0(m_{\sq_i}^2, m_{\sg}^2, m_q^2)\right] . \eeq

The one-loop corrected decay width in the \drbar scheme is then given by \beq
\Gamma(H^+\rightarrow\st_i\bar{\sb}_j)
=\frac{N_C\kappa_{\rm pole}}{16\pi m_H^3}[|G_{ij}|^2 +2G_{ij}{\rm
Re}(\delta G_{ij}^{(v)}
+\delta G_{ij}^{(w)})]. \label{20} \eeq
Here $\kappa_{\rm pole}$ refers to $\kappa$ in (6) evaluated with pole
squark masses.
The width of (20) is infrared divergent.

In the numerical analysis we take the pole quark masses as inputs. The
\drbar quark
masses are obtained from the pole quark masses by using \bea
m_q(Q)_{\drbar}&=&m_q({\rm pole})-\frac{\alpha_sC_F}{4\pi}
\left[ 2m_q(B_0(m_q^2, 0, m_q^2)-B_1(m_q^2, 0, m_q^2)) \right. \nonumber\\
&&+\sin 2\theta_\sq m_{\sg}(B_0(m_q^2, m_{\sg}^2, m_{\sq_1}^2)
-B_0(m_q^2, m_{\sg}^2, m_{\sq_2}^2))\nonumber\\ &&
\left. +m_q(B_1(m_q^2, m_{\sg}^2, m_{\sq_1}^2) +B_1(m_q^2, m_{\sg}^2,
m_{\sq_2}^2))\right] , \eea
which is derived from the graphs of Fig.~1d. Furthermore, in the phase
space term
$\kappa_{\rm pole}$ in (20) we have to take the pole squark masses given by
\bea
m_{\sq_i}^2({\rm
pole})&=&m_{\sq_i}^2(Q)_{\drbar}-\Pi_{ii}^{\sq}(m_{\sq_i}^2) \nonumber \\
&=&m_{\sq_i}^2(Q)_{\drbar}+\frac{\alpha_sC_F}{4\pi}
\left[ -3A(0)+4m_{\sq_i}^2B_0(m_{\sq_i}^2, 0, m_{\sq_i}^2)
+2m_{\sq_i}^2B_1(m_{\sq_i}^2, 0, m_{\sq_i}^2) \right. \nonumber \\
&&-\cos^22\theta_\sq A(m_{\sq_i}^2)-\sin^22\theta_\sq
A(m_{\sq_{i'}}^2)+4A(m_q^2)
\nonumber\\
&&+4m_{\sg}^2B_0(m_{\sq_i}^2, m_{\sg}^2, m_q^2)
+4m_{\sq_i}^2B_1(m_{\sq_i}^2, m_{\sg}^2, m_q^2) \nonumber\\
&&\left. -(-)^{i-1}4\sin 2\theta_\sq m_qm_{\sg} B_0(m_{\sq_i}^2, m_{\sg}^2,
m_q^2) \right] .
\label{22} \eea

\section{Gluon emission}
The infrared divergences in (\ref{20}) are cancelled by including the
${\cal O}(\alpha_s)$
contribution from real gluon emission from $\st$ and $\bar{\sb}$ (Fig.~1e).
The decay width of $H^+(p)\rightarrow\st_i(k_1)+\bar{\sb}_j(k_2)+g(k_3)$ is
given
in terms of the \drbar parameters as \beq
\Gamma(H^+\rightarrow\st_i\bar{\sb}_jg)
=\frac{\alpha_sC_FN_C|G_{ij}|^2}{4\pi^2 m_H}
[(m_H^2-m_{\st_i}^2-m_{\sb_j}^2)I_{12}
-m_{\st_i}^2I_{11}-m_{\sb_j}^2I_{22}-I_1-I_2]. \label{23} \eeq
The functions $I_n$, and $I_{nm}$ are defined as \cite{8} \beq
I_{i_1\ldots i_n}=\frac{1}{\pi^2}
\int\frac{d^3k_1}{2E_1}\frac{d^3k_2}{2E_2}\frac{d^3k_3}{2E_3}
\delta^4(p-k_1-k_2-k_3)\frac{1}
{(2k_3k_{i_1}+\lambda^2)\ldots(2k_3k_{i_n}+\lambda^2)}. \eeq
The explicit forms of $I_{i_1\ldots i_n}$ are given in \cite{8}. In (\ref{23}),
$I_{11,22,12}$ are infrared divergent. We have checked that the infrared
divergences in
(\ref{23}) cancel those in (\ref{20}). In the numerical analysis we define
the corrected
decay width as
$\Gamma^{\rm corr}(H^+\rightarrow\st_i\bar{\sb}_j)\equiv
\Gamma(H^+\rightarrow\st_i\bar{\sb}_j)+
\Gamma(H^+\rightarrow\st_i\bar{\sb}_jg)$.

\section{Numerical results and conclusions} As in ref.\cite{5}, we choose
\{ $m_H$, $m_{t,b}$(pole), $M$, $\mu$, $\tan\beta$, $M_{\tilde{Q}}$, $A$ \}
as the basic input parameters of the MSSM, taking
$M=(\alpha_2/\alpha_s)m_{\sg}=(3/5\tan^2\theta_W)M'$,
$M_{\tilde{Q}}\equiv M_{\tilde{Q}}(Q)_{\drbar}=
M_{\tilde{U}}(Q)_{\drbar}=M_{\tilde{D}}(Q)_{\drbar}=M_{\tilde{L}}$ and
$A\equiv A_t(Q)_{\drbar}=A_b(Q)_{\drbar}=A_{\tau}$. Here $M$ ($M'$) is
the SU(2) (U(1)) gaugino mass, $\alpha_2=g^2/4\pi$, and ($M_{\tilde{L}}$,
$A_{\tau}$)
are the mass matrix parameters of the slepton sector \cite{5}.
The parameters $M$, $M'$, $M_{\tilde{L}}$, and $A_{\tau}$ do not receive
${\cal O}(\alpha_s)$ QCD corrections.
The theoretical and experimental constraints for the basic input
parameters are described in ref.\cite{5}.
We take $m_Z=91.2$GeV, $m_W=80$GeV,
$m_t({\rm pole})=180$GeV \cite{mtop}, $m_b({\rm pole})=5$GeV,
$\sin^2\theta_W=0.23$
and $\alpha_2=\alpha_2(m_Z)=\alpha/\sin^2\theta_W =(1/129)/0.23=0.0337$.
For the running QCD coupling at the renormalization scale $Q$,
$\alpha_s=\alpha_s(Q)$,
we always take the one-loop expression $\alpha_s(Q)=
12\pi/\{(33-2n_f)\ln(Q^2/\Lambda_{n_f}^2)\}$,
with $\alpha_s(m_Z)=0.12$, and the number of quark flavors $n_f=5(6)$ for
$m_b<Q\le m_t$ (for $Q>m_t$).

We define the QCD corrections as the difference between the ${\cal
O}(\alpha_s)$ corrected
width $\Gamma^{\rm corr}_{ij}\equiv \Gamma^{\rm
corr}(H^+\rightarrow\st_i\bar{\sb}_j)$
of eqs.(20) plus (23) and the
tree-level width $\Gamma_{ij}^{\rm tree}\equiv
\Gamma^{(0)}(H^+\rightarrow\st_i\bar{\sb}_j)$
of eq.(6) where ($m_q(\rm pole)$, $M_{\tilde{Q}}$, $A$) are substituted for
($m_q$, $M_{\tilde{Q},\tilde{U},\tilde{D}}$, $A_q$). The QCD corrections depend
on the renormalization scale $Q$. We choose the optimum value of $Q$
($Q_{\rm opt}$)
such that $\displaystyle\Delta(Q)\equiv\sum_{q=t,b}\sum_{i=1,2}
(m_{\sq_i}({\rm tree})-m_{\sq_i}({\rm pole}))^2$ is minimized, where
$m_{\sq_i}(\rm tree)$ refers to
the $\sq_i$ mass defined by (1--4) calculated with $m_q(\rm pole)$,
$M_{\tilde{Q}}$, and $A$.

In order not to vary too many parameters, in the following we fix $\mu=300$GeV,
and take the values of $M$ and $\tan\beta$ such that
$m_{\tilde{\chi}_1^0}\simeq 50$GeV
as in \cite{5} where $\tilde{\chi}_1^0$ is the lightest neutralino.
In Fig.2 we show the $m_H$ dependence of the tree-level and corrected widths
$\displaystyle\Gamma^{\rm tree}(\st\bar{\sb})\equiv
\sum_{i,j=1,2}\Gamma^{\rm tree}_{ij}$ and
$\displaystyle\Gamma^{\rm corr}(\st\bar{\sb})\equiv
\sum_{i,j=1,2}\Gamma^{\rm corr}_{ij}$, and
the tree-level branching ratio $\displaystyle B^{\rm tree}(\st\bar{\sb})\equiv
\sum_{i,j=1,2}B^{\rm tree}(H^+\rightarrow\st_i\bar{\sb}_j)$ \cite{5} for (a)
$M_{\tilde{Q}}=250$GeV, $A=650$GeV, $\tan\beta=2$, $M=120$GeV, and (b)
$M_{\tilde{Q}}=136$GeV, $A=260$GeV, $\tan\beta=12$, $M=110$GeV.
In these two cases we have (in GeV units): (a) $Q_{\rm opt}=216.5$,
$m_{\sg}=380$, $(m_{\st_1},m_{\st_2}, m_{\sb_1},m_{\sb_2})\{(\rm
tree),(Q_{\rm opt})_{\drbar},
(\rm pole)\}=$ $\{(60,429,251,254)$, $(70,420,251,254)$, $(58,428,258,262)\}$,
$m_{\tilde{\chi}_1^+}=94$, and
(b) $Q_{\rm opt}=213.5$, $m_{\sg}=349$, $(m_{\st_1},m_{\st_2},
m_{\sb_1},m_{\sb_2})\{(\rm tree), (Q_{\rm opt})_{\drbar},(\rm pole)\}=$
$\{(81,302,62,193)$, $(74,292,120,164)$, $(66,302,126,171)\}$,
$m_{\tilde{\chi}_1^+}=98$.
Here $\tilde{\chi}_1^+$ is the lighter chargino. In both cases we see that the
$\st\bar{\sb}$ mode dominates the $H^+$ decay in a wide $m_{H^+}$ range at the
tree level, and that the QCD corrections to the $\st\bar{\sb}$ mode are
significant,
but that as a whole they do not invalidate the $\st\bar{\sb}$ mode dominance.

In Table 1 we show the values of the $B^{\rm tree}(\st\bar{\sb})$, the QCD
corrections
$C\equiv(\Gamma^{\rm corr}(\st\bar{\sb})- \Gamma^{\rm tree}(\st\bar{\sb}))
/\Gamma^{\rm tree}(\st\bar{\sb})$, and $C_{ij}\equiv(\Gamma^{\rm corr}_{ij}
-\Gamma^{\rm tree}_{ij})/ \Gamma^{\rm tree}_{ij}$ for
typical values of $M_{\tilde{Q}}$ and $A$, for (a)
$m_{H^+}=400$GeV, $\tan\beta=2$, $M=120$GeV, and
(b) $m_{H^+}=400$GeV, $\tan\beta=12$, $M=110$GeV.
We see again that the $\st\bar{\sb}$ mode dominates in a wide region also
when the QCD corrections are included. The QCD corrections can be very large
at some points of $(M_{\tilde{Q}},A)$; e.g. $C=-0.692$ at (175GeV, 0GeV) and
$C_{12}=0.734$ at (225GeV, $-350$GeV) in Table 1b. This occurs when
$m_H\sim m_{\st_i}+m_{\sb_j}$.
This enhancement is just a kinematical effect due to the QCD corrections to
$m_{\st_i}$ and $m_{\sb_j}$.

In conclusion, we have calculated the ${\cal O}(\alpha_s)$ QCD corrections
to the
decay width of $H^+\rightarrow\st_i\bar{\sb}_j$, including all quark mass
terms and $\sq_L-\sq_R$ mixing.
We find that the QCD corrections are significant but that they do not
invalidate our
previous conclusion at tree-level about the dominance of the $\st\bar{\sb}$
mode in
a wide MSSM parameter region.

\section*{Acknowledgements}

The work of Y.Y. was supported in part by the Fellowships of the Japan
Society for the Promotion of
Scienceand the Grant-in-Aid for Scientific
Research from the Ministry of Education, Science and Culture of Japan, No.
06-1923 and 07-1923.
The work of A.B., H.E., and W.M. was supported by the ``Fonds zur
F\"orderung der
wissenschaftlichen Forschung'' of Austria, project no. P10843-PHY.
The authors are grateful to Y. Kizukuri for the collaboration at the early
stage of this work.

\newpage

\section*{\large Table 1}

$B^{\rm tree}(\st\bar{\sb})$, $C$, and $C_{ij}$ for typical values of
$M_{\tilde{Q}}$ and
$A$, for ($m_H$(GeV), $m_t(\rm pole)$(GeV), $M$(GeV), $\mu$(GeV),
$\tan\beta$)$=$
(400, 180, 120, 300, 2) (a) and (400, 180, 110, 300, 12) (b).
The LEP bounds $m_{\sq,\tilde{l},\tilde{\nu}}
\rlap{\lower 3.5 pt \hbox{$\mathchar\sim$}}\raise 1pt \hbox{$>$} 45$GeV and
the requirement $m_{\st_1,\sb_1,\tilde{l}}>m_{\tilde{\chi}_1^0} (\simeq 50$GeV)
are satisfied. $Q_{\rm opt}$ denotes the optimum value of the
renormalization scale
$Q$ as defined in the text. The $*$ in the column of $C_{ij}$ means that the
$\st_i\bar{\sb}_j$ channel is not open at the tree level and/or one-loop level.
In Table 1b the columns of $C_{21}$ and $C_{22}$ are omitted, because the
corresponding channels are not open, except $C_{21}=-0.141$ at
$(M_{\tilde{Q}}, A)
=(150{\rm GeV}, 0{\rm GeV})$.

\newpage

(a)

\vspace{0.5cm}
\hspace{-1.8cm}
\begin{tabular} {cc|cc|cccc|c}
\hline \vphantom{$\frac{\frac{A^X}{B}}{C}$}
$M_{\tilde{Q}}$(GeV) & $A$(GeV) & $B^{\rm tree}(\st\bar{\sb})$ &
$C$ & $C_{11}$ & $C_{12}$ & $C_{21}$ & $C_{22}$ & $Q_{\rm opt}$(GeV) \\ \hline
75 & 0 & 0.715 & 0.054 & --0.251 & 0.143 & 0.025 & 0.664 & 224.0 \\
75 & 200 & 0.793 & 0.158 & --0.090 & 0.363 & --0.211 & 0.319 & 225.0 \\
75 & 300 & 0.818 & 0.161 & --0.181 & 0.261 & --0.229 & 0.363 & 240.5 \\
100 & 0 & 0.690 & 0.090 & --0.229 & 0.165 & 0.115 & 1.051 & 222.0 \\
100 & 300 & 0.788 & 0.219 & --0.166 & 0.295 & --0.181 & 0.509 & 234.5 \\
125 & 100 & 0.689 & 0.240 & --0.170 & 0.332 & 0.127 & 0.891 & 217.5 \\
125 & 350 & 0.702 & 0.175 & --0.220 & 0.267 & $*$ & $*$ & 238.0 \\
150 & 300 & 0.607 & 0.280 & --0.124 & 0.402 & $*$ &	$*$ & 221.0 \\
150 & 400 & 0.719 & 0.159 & --0.272 & 0.232 & $*$ &	$*$ & 238.0 \\
200 & 500 & 0.730 & 0.188 & --0.375 & 0.215 & $*$ &	$*$ & 226.5 \\
250 & 650 & 0.752 & 0.209 & --0.437 & 0.236 & $*$ &	$*$ & 216.5 \\
\hline
\end{tabular}
\vspace{0.5cm}

(b)

\vspace{0.5cm}
\hspace{-1.8cm}
\begin{tabular} {cc|cc|cc|c}
\hline \vphantom{$\frac{\frac{A^X}{B}}{C}$}
$M_{\tilde{Q}}$(GeV) & $A$(GeV) & $B^{\rm tree}(\st\bar{\sb})$ & $C$ & $C_{11}$
& $C_{12}$ & $Q_{\rm opt}$(GeV) \\
\hline
150 & --200 & 0.736 & --0.152 & --0.200 & 0.531 & 157.5 \\
150 &	0 & 0.715 & 0.062 & --0.383 & $*$ & 159.0 \\
150 & 250 & 0.740 & --0.013 & 0.047 & 0.394 & 200.0 \\
175 & --250 & 0.651 & 0.120 & --0.132 & 0.399 & 167.5 \\
175 &	0 & 0.612 & --0.692 & --0.432 & $*$ & 169.0 \\
175 & 300 & 0.681 & 0.251 & 0.139 & 0.323 & 204.5 \\
200 & --300 & 0.606 & 0.169 & --0.061 & 0.395 & 166.5 \\
200 & 350 & 0.656 & 0.306 & 0.239 & 0.345 & 201.0 \\
225 & --350 & 0.512 & 0.352 & 0.012 & 0.734 & 161.5 \\
225 & 400 & 0.588 & 0.501 & 0.355 & 0.587 & 193.5 \\
250 & 500 & 0.616 & 0.383 & 0.579 & 0.319 & 193.5 \\ \hline
\end{tabular}
\newpage

\vspace{20mm}
\section*{Figure Captions}
\renewcommand{\labelenumi}{Fig.\arabic{enumi}} \begin{enumerate}

\vspace{6mm}
\item
All diagrams relevant to the calculation of the ${\cal O}(\alpha_s)$
SUSY QCD corrections to the width of $H^+\rightarrow\st_i\bar{\sb}_j$ in
the MSSM.

\vspace{6mm}
\item
The $m_H$ dependence of $\Gamma^{\rm tree}(\sum\st_i\bar{\sb}_j)
\equiv\Gamma^{\rm tree}(\st\bar{\sb})$ (dashed line),
$\Gamma^{\rm corr}(\sum\st_i\bar{\sb}_j) \equiv\Gamma^{\rm
corr}(\st\bar{\sb})$ (solid line),
and $B^{\rm tree}(\st\bar{\sb})$ (short-dashed line) for ($m_t(\rm pole)$(GeV),
$M$(GeV), $\mu$(GeV), $\tan\beta$, $M_{\tilde{Q}}$(GeV),
$A$(GeV))$=$(180, 120, 300, 2, 250, 650)(a) and (180, 110, 300, 12, 136,
260)(b).
$\Gamma^{\rm tree}(\st_i\bar{\sb}_j)\equiv\Gamma^{\rm tree}_{ij}$ (dashed
lines) and
$\Gamma^{\rm corr}(\st_i\bar{\sb}_j)\equiv\Gamma^{\rm corr}_{ij}$ (solid
lines) are
separately shown for $(i,j)=$(1,1) and (1,2).

\end{enumerate}

%Figure 1: PAGE with Feynman graphs:
\clearpage

\begin{center}
\begin{picture}(412,556)
\put(0,0){\mbox{\psfig{file=fig1.eps}}} \setlength{\unitlength}{1mm}
%Figure 1a:
\put(74,156){\makebox(0,0)[t]{\large{\bf{Fig.~1a}}}}
\put(55,180){\makebox(0,0)[bl]{$H^+$}}
\put(95,197){\makebox(0,0)[l]{$\st_i$}}
\put(95,161){\makebox(0,0)[l]{$\bar{\sb}_j$}}

%Figures 1b:
\put(74,101){\makebox(0,0)[t]{\large{\bf{Fig.~1b}}}}
\put(3,126){\makebox(0,0)[bl]{$H^+$}}
\put(43.5,143){\makebox(0,0)[l]{$\st_i$}}
\put(29,132){\makebox(0,0)[b]{$\st_i$}}
\put(29,118){\makebox(0,0)[t]{$\bar{\sb}_j$}}
\put(37,125){\makebox(0,0)[l]{$g$}}
\put(43.5,107){\makebox(0,0)[l]{$\bar{\sb}_j$}}

\put(55,126){\makebox(0,0)[bl]{$H^+$}}
\put(81,132){\makebox(0,0)[b]{$t$}}
\put(81,118){\makebox(0,0)[t]{$b$}}
\put(87,125){\makebox(0,0)[l]{$\sg$}}
\put(95,143){\makebox(0,0)[l]{$\st_i$}}
\put(95,107){\makebox(0,0)[l]{$\bar{\sb}_j$}}

\put(106.5,126){\makebox(0,0)[bl]{$H^+$}}
\put(122.5,133.5){\makebox(0,0)[b]{$\st$}}
\put(122.5,116.5){\makebox(0,0)[t]{$\bar{\sb}$}}
\put(147,143){\makebox(0,0)[l]{$\st_i$}}
\put(147,107){\makebox(0,0)[l]{$\bar{\sb}_j$}}

%Figures 1c:
\put(73,56){\makebox(0,0)[t]{\large{\bf{Fig.~1c}}}}
\put(2,66.5){\makebox(0,0)[tl]{$\sq_i$}}
\put(14.5,66.5){\makebox(0,0)[t]{$\sq_i$}}
\put(14.5,77.5){\makebox(0,0)[b]{$g$}}
\put(27,66.5){\makebox(0,0)[tr]{$\sq_i$}}

\put(41,66.5){\makebox(0,0)[tl]{$\sq_i$}}
\put(53.5,83.5){\makebox(0,0)[b]{$g$}}
\put(66,66.5){\makebox(0,0)[tr]{$\sq_i$}}

\put(80,66.5){\makebox(0,0)[tl]{$\sq_j$}}
\put(92.5,66.5){\makebox(0,0)[t]{$q$}}
\put(92.5,76.5){\makebox(0,0)[b]{$\sg$}}
\put(105,66.5){\makebox(0,0)[tr]{$\sq_i$}}

\put(119,66.5){\makebox(0,0)[tl]{$\sq_j$}}
\put(131.5,82.5){\makebox(0,0)[b]{$\sq$}}
\put(144,66.5){\makebox(0,0)[tr]{$\sq_i$}}

%Figures 1d:
\put(21,-7){\makebox(0,0)[t]{\large{\bf{Fig.~1d}}}}
\put(8.5,22){\makebox(0,0)[tl]{$q$}}
\put(21,22){\makebox(0,0)[t]{$q$}}
\put(21,32.2){\makebox(0,0)[b]{$g$}}
\put(33.5,22){\makebox(0,0)[tr]{$q$}}

\put(8.5,3.1){\makebox(0,0)[tl]{$q$}}
\put(21,3.1){\makebox(0,0)[t]{$\sq$}}
\put(21,12.3){\makebox(0,0)[b]{$\sg$}}
\put(33.5,3.1){\makebox(0,0)[tr]{$q$}}

%Figures 1e:
\put(99.8,-7){\makebox(0,0)[t]{\large{\bf{Fig.~1e}}}}
\put(54.8,18.7){\makebox(0,0)[bl]{$H^+$}}
\put(94.8,35.7){\makebox(0,0)[l]{$\st_i$}}
\put(94.8,26){\makebox(0,0)[l]{$g$}}
\put(94.8,-0.3){\makebox(0,0)[l]{$\bar{\sb}_j$}}

\put(105.8,18.7){\makebox(0,0)[bl]{$H^+$}}
\put(146.8,35.7){\makebox(0,0)[l]{$\st_i$}}
\put(146.8,9){\makebox(0,0)[l]{$g$}}
\put(146.8,-0.3){\makebox(0,0)[l]{$\bar{\sb}_j$}} \end{picture}
\end{center}

%Figure 2: PAGE with graphs including widths and BR's \clearpage

\setlength{\unitlength}{1mm}
\begin{center}
\begin{picture}(150,215)
%\put(0,0){\framebox(150,215){}}
\put(0,25){\mbox{\psfig{file=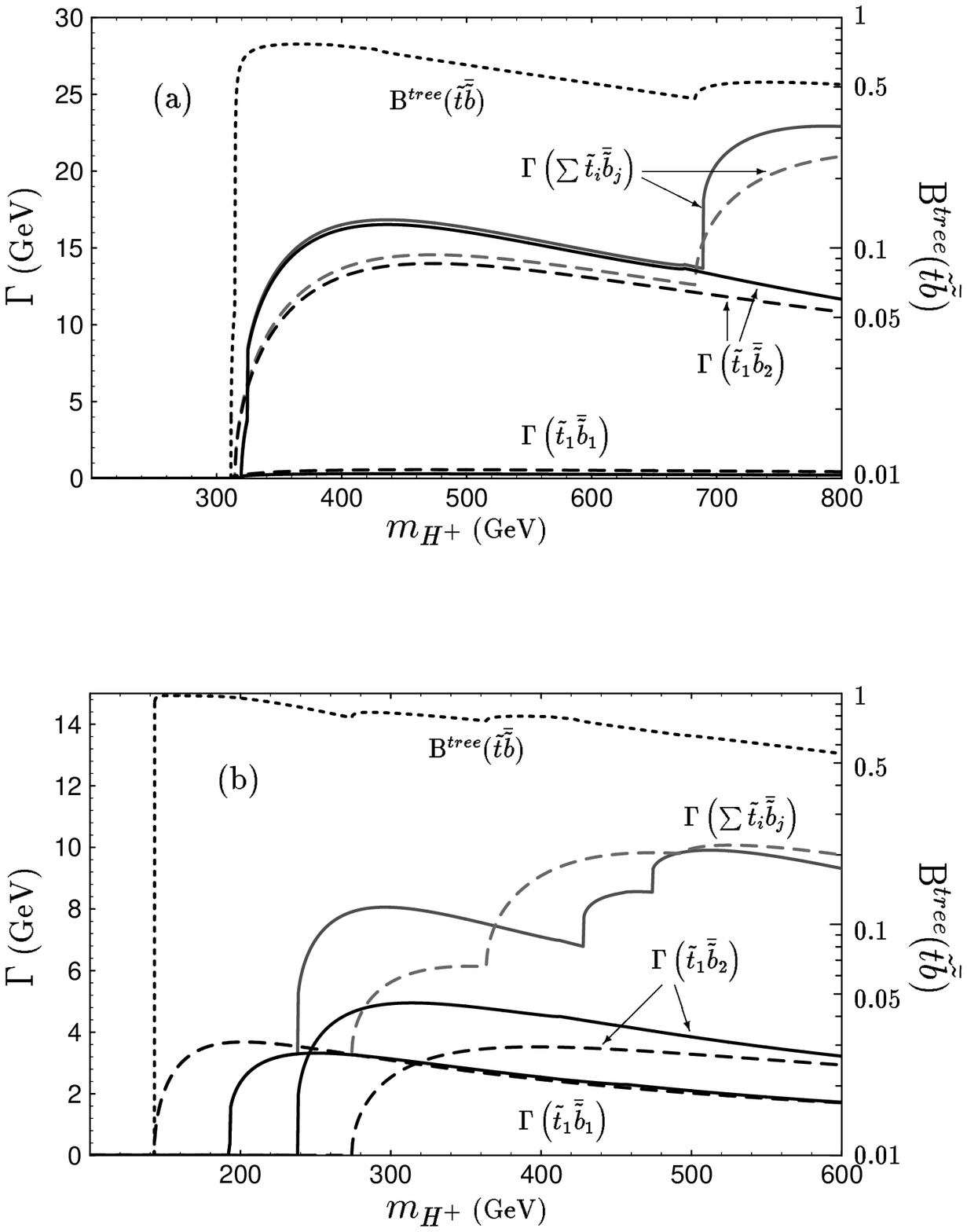,width=15cm}}}
\put(75,0){\makebox(0,0)[b]
{\Large \bf Fig. 2}} \end{picture}
\end{center}
\setlength{\unitlength}{1pt}

%%%%% End of document
\end{document}